\documentclass[default,iicol]{sn-jnl}

\usepackage{graphicx}
\usepackage[frozencache=true,cachedir=minted-cache]{minted}  
\usepackage[utf8]{inputenc}
\usepackage{subfig}
\usepackage{booktabs}
\usepackage{amsfonts}
\usepackage{siunitx}
\usepackage{amsmath}

\usepackage{lipsum}  
\usepackage[labelfont=bf]{caption}

\usepackage{hyperref}
\usepackage{url}
\usepackage{float}

\makeatletter

\newcommand\floatc@simplerule[2]{{\@fs@cfont #1 #2}\par}
\newcommand\fs@simplerule{\def\@fs@cfont{\bfseries}\let\@fs@capt\floatc@simplerule
  \def\@fs@pre{\hrule height.8pt depth0pt \kern4pt}%
  \def\@fs@post{\kern4pt\hrule height.8pt depth0pt \kern4pt \relax}%
  \def\@fs@mid{\kern8pt}%
  \let\@fs@iftopcapt\iftrue}

\makeatother

\floatstyle{simplerule}
\newfloat{infobox}{t}{lop}
\floatname{infobox}{Box}

\jyear{2022}%

\theoremstyle{thmstyleone}%
\usepackage[numbers,sort&compress]{natbib}
\theoremstyle{thmstyletwo}%

\theoremstyle{thmstylethree}%

\begin{document}

\title[Article Title]{Cait: analysis toolkit for cryogenic particle detectors in Python}


\author*[1]{\fnm{Felix} \sur{Wagner}}\email{felix.wagner@oeaw.ac.at}

\author[1,2]{\fnm{Daniel} \sur{Bartolot}}\email{daniel.bartolot@oeaw.ac.at}

\author[1,2]{\fnm{Damir} \sur{Rizvanovic}}\email{damir.rizvanovic@oeaw.ac.at}

\author[1,2]{\fnm{Florian} \sur{Reindl}}\email{florian.reindl@oeaw.ac.at}

\author[1,2]{\fnm{Jochen} \sur{Schieck}}\email{jochen.schieck@oeaw.ac.at}

\author[1]{\fnm{Wolfgang} \sur{Waltenberger}}\email{wolfgang.waltenberger@oeaw.ac.at}

\affil[1]{\orgdiv{Institut f{\"u}r Hochenergiephysik}, \orgname{{\"O}sterreichischen Akademie der Wissenschaften}, \orgaddress{\street{Nikolsdorfer Gasse 18}, \city{Wien}, \postcode{1050}, \state{Wien}, \country{{\"O}sterreich}}}

\affil[2]{\orgdiv{Atominstitut}, \orgname{Technische Universit{\"a}t Wien}, \orgaddress{\street{Stadionallee 2}, \city{Wien}, \postcode{1020}, \state{Wien}, \country{{\"O}sterreich}}}


\abstract{Cryogenic solid state detectors are widely used in dark matter and neutrino experiments, and require a sensible raw data analysis. For this purpose, we present Cait, an open source Python package with all essential methods for the analysis of detector modules fully integrable with the Python ecosystem for scientific computing and machine learning. It comes with methods for triggering of events from continuously sampled streams, identification of particle recoils and artifacts in a low signal-to-noise ratio environment, the reconstruction of deposited energies, and the simulation of a variety of typical event types. Furthermore, by connecting Cait with existing machine learning frameworks we introduce novel methods for better automation in data cleaning and background rejection.}




\maketitle


\section{Introduction}\label{sec1}

Rare Event Searches (RES) in particle physics require the sensitive measurement of tiny energy depositions from scattering of dark matter particles or coherent elastic neutrino-nucleus scattering \cite{PhysRevD.31.3059, akimov_observation_2017}. Multiple detector technologies can be realised, among them cryogenic solid state detectors, equipped with Transition Edge Sensors (TESs) \cite{probst_model_1995}, which feature strong sensitivity for sub-keV recoil energies. At operation temperatures of $\mathcal{O}$(mK), a low energy particle recoil inside the target produces a distinct athermal phonon population. The population is collected by the TES and the resulting change in the electronic resistance is transformed into a voltage signal using a Superconducting QUantum Interference Device (SQUID) amplifier. The Raw Data (RD), i.e. the sensor signal, consist therefore of one or multiple time series of voltage traces, depending on the number of sensors used simultaneously in the measurement. Careful RD preparation is crucial to acquire reliable physics results. This includes the triggering of events from aforementioned continuous data streams, data cleaning and background rejection, calibration and reconstruction of recoil energies and the estimation of resolutions and efficiencies. The individual challenges of these tasks vary with the measurement conditions: low-threshold experiments require high Signal-to-Noise Ratio (SNR); underground physics runs demand high efficiency at low rates; and above-ground runs must handle high rates with low purity. 

Many of the steps in the RD analysis are common signal processing tasks within the broader scientific computing and data science fields: extraction of time series features, regression and classification problems, or data visualization. In the past decade, software packages emerged that enable these tasks to be done fully in Python or with Python APIs \cite{10.5555/1593511}. 
This comes with various advantages such as Python's ease of use, the ability to work both with scripted pipelines and in interactive sessions, and the large number of open source Python modules that are publicly available. Therefore, the Python ecosystem is the preferred choice of many users for data analysis. To make this ecosystem also accessible in our RD analysis, we developed Cryogenic Artificial Intelligence Tools (Cait), a package written fully in Python, depending only on standard scientific computing Python packages. Cait provides a framework for the management of raw as well as processed data from cryogenic detectors in structured files, and a variety of tools to handle all necessary tasks of the analysis. The workflow requires a minimum of manual user interaction and is therefore scalable to multi-detector setups. 

The rapid development in the fields of big data and machine learning in the past years motivate new methods for data analysis, beyond a traditional handcrafted, cut-based analysis. Methods for a machine learning-based analysis of direct detection dark matter experiments were already proposed in the literature, e.g. in Ref. \cite{Armengaud_2016, arnaud_first_2018, Coarasa_2021, golovatiuk_deep_2022, Khosa_2020, https://doi.org/10.48550/arxiv.2201.05734}. With Cait, we provide dedicated PyTorch data sets and PyTorch Lightning data modules \cite{NEURIPS2019_9015, falcon2019pytorch}, to apply supervised classification with neural networks directly to the raw or processed data. Furthermore, we implemented frameworks to simulate realistic data sets from cryogenic detectors, i.e. particle recoil signatures as well as detector artifacts and backgrounds. The combination of supervised machine learning with realistic, simulated data sets allows for the training of models for fully automatic and blind data cleaning and background rejection, which can reduce the needed personpower for data quality monitoring and raw data analysis significantly.

Cait was initially developed for the application of machine learning methods in the data analysis of the CRESST and COSINUS experiments \cite{PhysRevD.100.102002, https://doi.org/10.48550/arxiv.2111.00349}. It is an open source package under the GPL-3.0 license, available via the Python package index and on GitHub \cite{cait_wagner2020, pypi}. Release versions are hosted on Zenodo \cite{felix_wagner_2022_6359433}. This work is based on its 1.1 release and is structured as follows: 

In Sec.~\ref{sec: data} we introduce the workflow of the data analysis, starting from the triggering of data, and ending with the validation of the analysis chain with simulated data. We introduce the data organization in structured files and the integration with individual, Python-based methods. In Sec.~\ref{sec:examples} we highlight experiments with three features implemented in Cait: the simulation of labeled data sets with different classes of events, the training and testing of a neural network for RD cleaning, and a tool for the interactive visualization and event selection. In Sec.~\ref{sec:conclusion} we address potential applications and future developments.

With this work, we aim to make the following contributions to the scientific discussion. First, Cait can be of interest for all analysts of data from cryogenic detectors. Second, the implemented methods can as well serve as a template for individual implementations in similar open source projects. Finally, we want to join the ongoing efforts in the community towards a common culture of reproducibility and comparability of scientific results.

Cait depends on several publicly available packages for scientific computing (NumPy, SciPy, Numba, Pandas \cite{2020SciPy-NMeth, harris2020array, 10.1145/2833157.2833162, mckinney2010data}), for data visualization (Matplotlib, Plotly \cite{Hunter:2007, plotly}), for interactive workflows (IPython, ipywidgets \cite{PER-GRA:2007, interactive_Jupyter_widgets}) and for the plotting of progress bars (tqdm \cite{da_costa_luis_casper_o_2019_3435774}). 

\section{Data organization and analysis workflow} \label{sec: data}

\begin{figure*}[t]
\centering
\includegraphics[width=\linewidth]{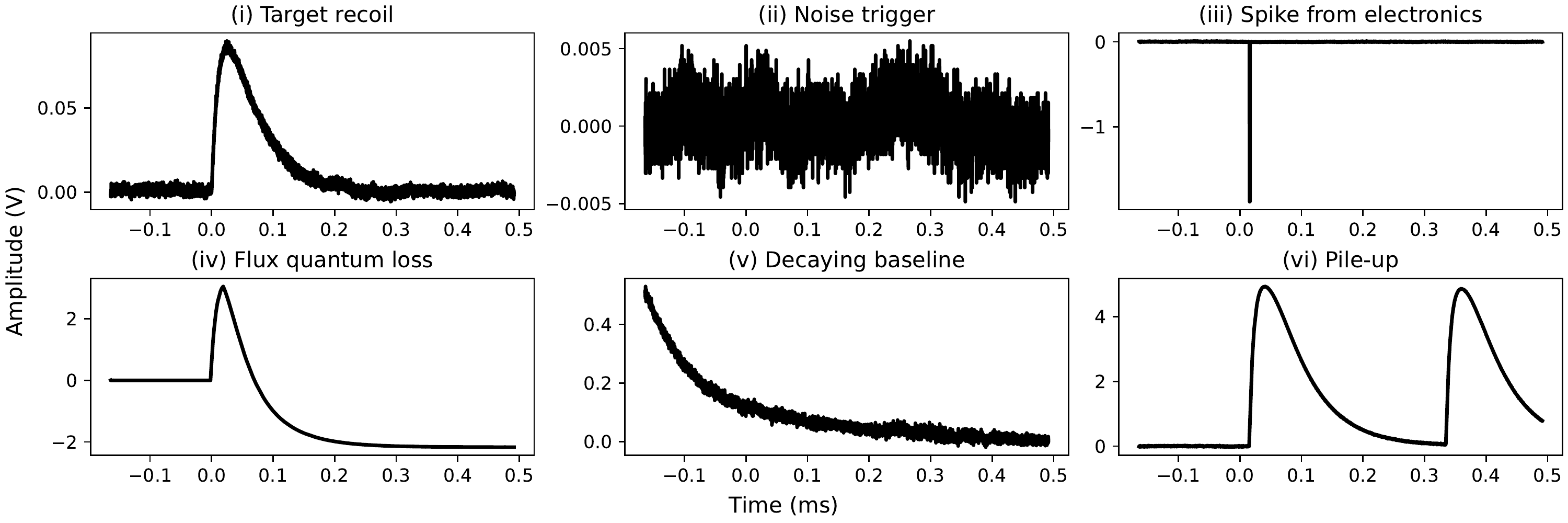}
\caption{Cait can simulate voltage traces of events. Apart from pulse shaped target recoils (i) and noise triggers (ii), a set of typical artifacts occur frequently in measured data. The electronics of the data acquisition can introduce spikes or glitches (iii). SQUID readout systems can transition between individual flux quanta, causing discrete jumps in the voltage offset (iv). Pulses can pile up in various ways, fully or partially within the same recording window, leading to strongly decaying voltage baselines (v) or multiple pulse shapes within one window (vi).}
\label{sim_events}
\end{figure*}

Cryogenic solid state detectors are used in experiments that require sensitivity to small recoil energies, i.e. a low detection threshold. In order to improve the sensitivity, two significant improvements have been made in recent years: the manufacturing of detector modules with smaller targets, which led to a larger collection efficiency for athermal phonons \cite{PhysRevD.96.022009}; and the recording of continuous data streams for offline triggering with optimized trigger methods. Addressing the former measure, even though a smaller target leads to increased sensitivity, it also decreases the amount of collected exposure in a given run time. Therefore, many experiments plan to scale up the number of detectors operated simultaneously \cite{Billard_2022, PhysRevD.95.082002, Undagoitia_2015}. Addressing the latter measure, an optimized threshold increases the need for sensible cleaning of triggered data despite a low SNR, as surviving artifacts can significantly impact physics results. These considerations lead to the following requirements on the analysis software:

\begin{itemize}
\item The ability to trigger a continuously recorded data stream with a trigger method, individually optimized for each channel.
\item Extraction of descriptive features from the triggered RD events to enable data cleaning and background rejection despite low SNR.
\item Automation of as many steps of the analysis as possible, to achieve scalability to a multi-detector setup.
\item An efficient and interactive workflow for those steps of the analysis where manual work is necessary.
\end{itemize}

Before we discuss how Cait meets these requirements, let us review the vocabulary which will be used throughout the following sections. We call an event any occurence that causes a sensor signal that is at least in one channel significant enough to trigger. In this work, we do not distinguish between the event and the signal that is caused by the event. Events can have different origins, and examples of simulated, typical events are shown in Fig.~\ref{sim_events}. We mainly distinguish events in target recoils and artifacts, which are any other type of energy deposition or detector effect that is triggered but should be removed in the process of data cleaning. In our classification, pile-up of target recoil signatures count as an artifact, as a meaningful reconstruction of the involved recoil energies is not possible with standard methods. The target recoils can be further classified as signal or background events, in a subsequent analysis. This classification depends on the type of sought-for physics. In a direct detection dark matter search, nuclear recoils would count as signal candidates, while electron recoils count as background. This classification is typically performed by the use of additional detectors (multi-channel readout) and parametric models of the expected signal and background rates. However, the classification in signal and background is not strictly part of the RD analysis, and therefore not subject of this work.

\begin{figure*}[!h]
\centering
\includegraphics[width=\linewidth]{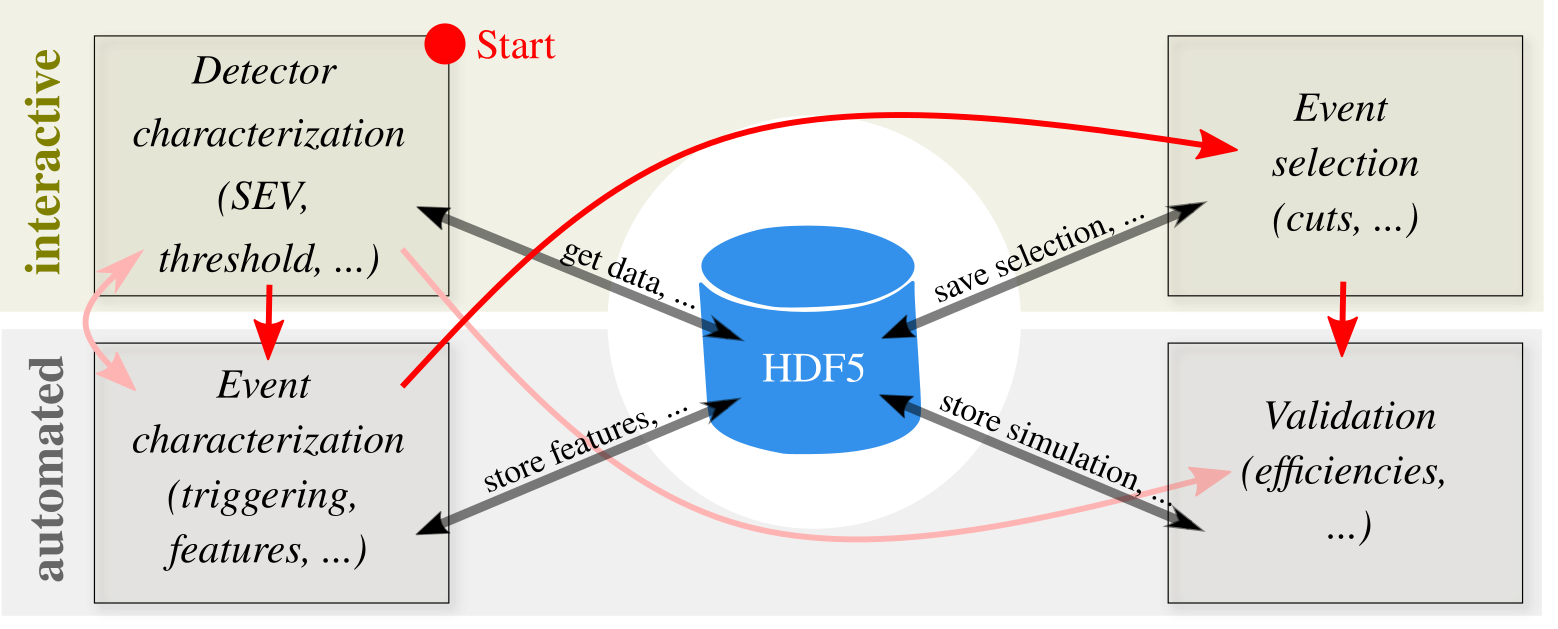}
\caption{Visualization of the four steps of the analysis workflow: detector characterization, event characterization, event selection, validation. The tasks which require significant run time are automated, i.e. do not require user interaction. The red arrows indicate the usual order of tasks in the analysis. However, the time intensive part of the validation (simulating events and calculating their features) can already be done in parallel with the event characterization. Also, the detector and event characterization has to go through at least two iterations: events need to be triggered to characterize the detector, and triggering with optimized threshold is only possible after the detector characterization.}
\label{flowchart}
\end{figure*}

In the following, we guide the reader through the major steps of the analysis workflow, and include minimal code examples of the discussed content. The workflow is visualized in Fig.~\ref{flowchart}.

\subsection{Triggering and the structured file format}

We consider the scenario in which an experiment has recorded the continuous data stream from a sensor. The data is stored as a chain of 16-bit precision samples. A mock stream of such data can be simulated with the code included in App. \ref{app:mockstream}, and we call the format in this work a "csmpl" file. The first step in the exploration of the data is the triggering of events from the stream, with a preliminary and typically conservatively chosen analysis threshold. For each trigger, a fixed number of samples surrounding the triggered sample are stored, forming record windows. We use the Hierarchical Data Format version 5 (HDF5, \cite{hdf5, collette_python_hdf5_2014}) to store the time series of triggered events , an open source file format that supports large, complex, heterogeneous data. The ``file system''-like structure allows to store the time series, calculated features and additional information about the measurement in a single file. We use a dedicated class, the \verb|DataHandler|, to write and read from the HDF5 file, and apply methods to data stored in the file. The code example below shows these steps: instantiation of a DataHandler object linked to an HDF5 file, triggering of a continuously recorded data stream, and storing the triggered time series in the HDF5 file.

\vspace{0.5cm}
\begin{minted}[
baselinestretch=1,
% linenos
]{python}
import cait as ai
import numpy as np

dh = ai.DataHandler(nmbr_channels=1, 
    record_length=16384, 
    sample_frequency=25000)
dh.set_filepath(path_h5='./', 
    fname='mock_001', 
    appendix=False)
                
dh.include_csmpl_triggers(
    csmpl_paths=['./mock_001.csmpl',],
    thresholds=[0.1,],
    of=[np.ones(8193)])

dh.include_triggered_events(
    csmpl_paths=['./mock_001.csmpl',],
    exclude_tp=False)
\end{minted}
\vspace{0.5cm}

The HDF5 file has a uniform structure: it contains folders (groups) to hold individual numerical arrays (data sets). General information about the measurements (start time, time base, ...) is stored in the group \verb|metainfo|. The data sets inside this group are scalars. Time series and features of triggered events are stored in the group \verb|events|. The data set named \verb|event| is a three dimensional array, containing the time series of the signals in all detectors (channels). The first index corresponds to the channel, the second index to the event number, the third to the sample inside the time series. The stored data can be accessed as NumPy array:

\vspace{0.5cm}
\begin{minted}[
baselinestretch=1,
% linenos
]{python}
dh.get("events", "event", 0, 123, 4845) 
\end{minted}
\vspace{0.5cm}

\noindent returns the 4846'th sample of channel 0 of the 124'th triggered event. All data sets in the \verb|events| group have the shape: \verb|(channels, events, features)|. In the case where the information of the data set is not specific to individual channels, the data set has only two dimensions, with the first dimension (the channel index) missing. In the case where it consists only of an individual feature, it has only two dimensions as well, with the third dimension (the feature index) missing. Some data sets, e.g. the elapsed time since start of the measurement (\verb|hours|), have only one dimension (the event index).

The size of the HDF5 files is determined by the length and trigger rate of the measurement and length of the record windows of triggered events. To give an example: consider an experiment that operates two detectors, triggers events with a rate of 1 per minute, a record length of 16384 samples and a measurement time of one month. The corresponding HDF5 file has a size of 
\begin{equation*}
\begin{split}
2 \text{ channels } \times 720 \text{ hours } \times 60 \text{ triggers per hour } \\ \times 16384 \text{ samples } \times 4 \text{ B } = 5.27 \text{GB},
\end{split}
\end{equation*}
\noindent if the triggered time series are stored as 32-bit float values. 



\subsection{Detector characterization} \label{sec:detchar}

We use the events extracted with the preliminary trigger threshold to explore the properties of our detector. This especially includes the calculation of a standard event (SEV), the normal pulse shape for a target event, and the noise power spectrum (NPS) for each channel individually. From this information, we can calculate a matched (optimal) filter kernel (OF), which is used to optimize the trigger condition for our continuous data stream \cite{alduino_low_2017, MANCUSO2019492}. The following code example shows the discussed steps performed with Cait, details are added as comments to the code. Note that for the creation of the SEV and the NPS we apply a data quality criterion, to avoid both from being spoiled by pile-up events. For the SEV, we use a peak search algorithm, while for the NPS we use the residuals of a cubic polynomial fit.

\vspace{0.5cm}
\begin{minted}[
baselinestretch=1,
% linenos
]{python}
# calculate pulse shape parameters
dh.calc_mp()
# apply peak detection algorithm
dh.calc_peakdet()

# calculate SEV from events with one peak
n_peaks = dh.get('events', 'nmbr_peaks')
cut = n_peaks == 1
dh.calc_sev(use_idx=cut.nonzero()[1])

# include random triggers from stream
dh.include_noise_triggers(nmbr=200, 
    start_end=[0.001, 1.999], 
    max_distance=7200)

# include record windows of noise events
dh.include_noise_events(
    csmpl_paths=['./mock_001.csmpl',])

# do fit to baselines for cleaning
dh.calc_bl_coefficients()

# calculate NPS and OF
dh.calc_nps(percentile=0.5)
dh.calc_of()
\end{minted}
\vspace{0.5cm}

A low threshold analysis requires an optimized trigger threshold value. This value is often chosen as 5-7 times the baseline resolution, i.e. the standard deviation of the detector noise. The detector resolution can be estimated with randomly induced noise triggers. In the detector characterization we also want to establish an energy calibration, i.e. a translation of the recorded voltage values to recoil energies. To this end, one or multiple prominent features in the spectrum with known energy, often peaks from dedicated mono-energetic sources, are necessary. From them we can determine a conversion factor or function to map voltage values to their corresponding energies. For both of these tasks, dedicated methods are implemented in Cait. 

\subsection{Event characterization}
\label{sec:evchar}

We can use our knowledge about the detector for the characterization of the individual events in the measurement. For a low threshold analysis, the first step is to repeat the triggering from the continuous data stream with optimized threshold after filtering with an OF. This step is followed by the calculation of time series features which describe the pulse shape of the events. To provide a code example, this is mostly done by the application of methods of the DataHandler class, results are automatically stored in the HDF5 file:

\vspace{0.5cm}
\begin{minted}[
baselinestretch=1,
% linenos
]{python}
dh.calc_additional_mp()
dh.apply_of()
dh.apply_array_fit()
\end{minted}
\vspace{0.5cm}

Depending on the calculated features and the amount of measurement time, this step is relatively time intensive. To provide an example: Triggering of 10~h measurement time with one channel takes $\mathcal{O}(1)$ minute on a single-CPU worker node. The run time for the calculation of basic features, such as pulse heights, rise and decay times or filtering of triggered events depends on the number of events, but is usually negligibly small compared to the time necessary for triggering. Other time consuming tasks are the application of fit models to individual events, such as pulse shape fits or a non-linear energy calibration. The time necessary for the full processing of a 10 h measurement file usually ranges from 5 to 25 minutes on a 1-CPU worker node. In RES, the typical run times of measurements range up to years. However, this time consuming part of the analysis can be performed without any user interaction, as all parameters were already fixed in the previous step of the detector characterization. In our workflow, it is therefore fully scripted and can be submitted and processed in parallel as a series of small batch jobs for different intervals of the measurement. 


\subsection{Event selection}

Having calculated all descriptive features of all events, we can define cuts to discriminate between target recoils and artifacts. The application of quality cuts is a part of the analysis which has to be treated with care. Excluding too many target events or keeping too many artifacts can lead to a wrong signal signature or exclusion of the sought-for physics process. Cuts are defined as rejection regions in the feature space of the triggered events. Traditionally these rejection regions are handcrafted, i.e. determined by looking at distributions of descriptive features and applying a hard cut along a suitable value. In Cait, we use the LogicalCut class to combine and manage cuts for a certain data set. In the example below, we apply cuts on the pulse height, decay time and mean value of the events:

\vspace{0.5cm}
\begin{minted}[
baselinestretch=1,
fontsize=\footnotesize
% linenos
]{python}
cut = ai.cuts.LogicalCut()
cut.add_condition(
    dh.get('events', 'pulse_height')[0] < 1.5)
cut.add_condition(
    dh.get('events', 'decay_time')[0] < 50)
cut.add_condition(
    dh.get('events', 'mean')[0] > 0)
\end{minted}
\vspace{0.5cm}

Suitable values for data quality cuts need to be optimized for detector modules individually. This is most conveniently done with interactive tools, where feature distributions can be scattered against each other, and time series of individual events can be located in the feature distributions. We introduce a powerful tool for that task in Sec.~\ref{sec:viztool}.

Owing to the ubiquity of powerful classification methods, many analysts have experimented with supervised machine learning to achieve a more precise and less biased event selection. This approach is discussed in more detail in Sec.~\ref{sec:universaleventselection}, together with its implementation in Cait.

\subsection{Validation} 

The steps of event characterization, and especially event selection, need to be validated, to account for a potential loss of signal candidate events in the analysis chain. This is done by estimating an energy dependent efficiency, i.e. the product of trigger and cut survival probability of particle recoil events. Our approach for the efficiency estimation is to add the normal pulse shape of target recoils with randomly sampled recoil energies on the record windows of randomly induced noise triggers. To these simulated events we apply the full analysis chain and thus determine their survival rate. Below we show the code corresponding to the simulation of such target recoils, and the calculation of the same time series features that were calculated for the triggered events in Sec.~\ref{sec:evchar}:

\vspace{0.5cm}
\begin{minted}[
baselinestretch=1,
% linenos
]{python}
dh.simulate_pulses('./sim_001.h5', 
    size_events=200)

dh_res = ai.DataHandler(nmbr_channels=1, 
    record_length=16384, 
    sample_frequency=25000)
dh_res.set_filepath(path_h5='./', 
    fname='sim_001', appendix=False)

dh_res.calc_peakdet()
dh_res.calc_mp()
dh_res.calc_additional_mp()
dh_res.apply_of()
dh_res.apply_array_fit()
\end{minted}
\vspace{0.5cm}

The final result of the raw data analysis is a compact data set of recoil events, fully characterized by their recoil energy and potentially additional information from veto detectors, as e.g. a light or ionization yield parameter. In addition to that, we obtain the energy dependent efficiency, describing the sensitivity of the detector. We thus create a data set with all necessary information for obtaining physics results.

\section{Highlighted features and experiments} \label{sec:examples}

The implemented features of Cait exceed the standard steps of the analysis workflow which were presented in the previous section. Therefore, we highlight in this section a specific series of use cases: first we simulate realistic data sets of target recoils and artifacts; then we use these data sets to train a supervised machine learning model for automated data cleaning; finally, we show one of the data sets in an interactive visualization tool. We use the simulated data from our first experiment to perform the other consecutive two. The code producing all experiments from this section is part of the Cait documentation.

\begin{infobox}[htbp!]
\begin{description}
\itemsep0em
\item[\textit{Event pulse}] A pulse-shaped target recoil (see Fig.~\ref{sim_events}, i). We simulate them with a parametric model from Ref. \cite{probst_model_1995}. 
\item[\textit{Noise}] An empty noise trace (see Fig.~\ref{sim_events}, ii). 
\item[\textit{Decaying baseline}] The decaying part of a particle recoil, which happened before the start of the trigger record window.
\item[\textit{Temperature rise}] A sudden rise in temperature, producing a linear upwards drift of the baseline. 
\item[\textit{Spike}] Glitches in the digitizer can cause a small number of consecutive samples to have a malicious value, effectively producing a upward- or downward facing spike in the sensor signal (see Fig.~\ref{sim_events}, iii).
\item[\textit{SQUID jump}] A fast rise of the thermometer temperature, often due to an high energetic particle recoil, can cause the jump from a higher flux quantum state of the SQUID amplifier to a lower one, i.e. the loss of a flux quantum (see Fig.~\ref{sim_events}, iv). 
\item[\textit{Reset}] After a certain number of flux quanta losses, the SQUID amplifier resets to a higher voltage baseline level. 
\item[\textit{Carrier event}] Depending on the design of the cryogenic detector, there might be additional thermal components, other than the target crystal, in which particle recoils can happen. An example is a carrier crystal, a small separate crystal on which the TES is evaporated.
\item[\textit{Tail event}] A pulse shaped event with an additional, slowly decaying component, usually caused by a feedback effect, e.g. the reabsorption of scintillation light.
\item[\textit{Decaying baseline with event pulse}] A decaying baseline event, which is coincident with a particle recoil (see Fig.~\ref{sim_events}, v).
\item[\textit{Pile up}] Multiple particle recoils happening inside the same record window, seen as partially overlapping pulse shapes (see Fig.~\ref{sim_events}, vi).
\item[\textit{Early or late trigger}] In some scenarios the maximum of the pulse can appear too early or too late inside the record window, e.g. when a time interval with blocked trigger overlaps with a subsequent event.
\end{description}
\caption{Event classes.}
\label{evtypes}
\end{infobox}

\subsection{Multiclass event simulation} \label{sec:simulation}


The training of a supervised machine learning model requires a decent amount of labeled training data. For acquiring a labeled data set, two approaches have been taken in the past: dedicated measurements with known high rate of signal-like and background-like events, e.g. by employing particle sources (e.g. in Ref. \cite{Coarasa_2021}), or Monte Carlo simulation of high-level features of signal and background events (e.g. in Ref. \cite{Armengaud_2016}). Contrary to these methods, we simulate in Cait an arbitrary amount of labeled time series corresponding to realistic target recoils and artifacts directly. Those can be subjected to the same analysis chain as triggered events from measured data. This way we can effectively sample from the feature distribution of typical event classes, which can also be useful for testing data quality cuts.  

\begin{figure}[t]
\centering
\includegraphics[width=\linewidth]{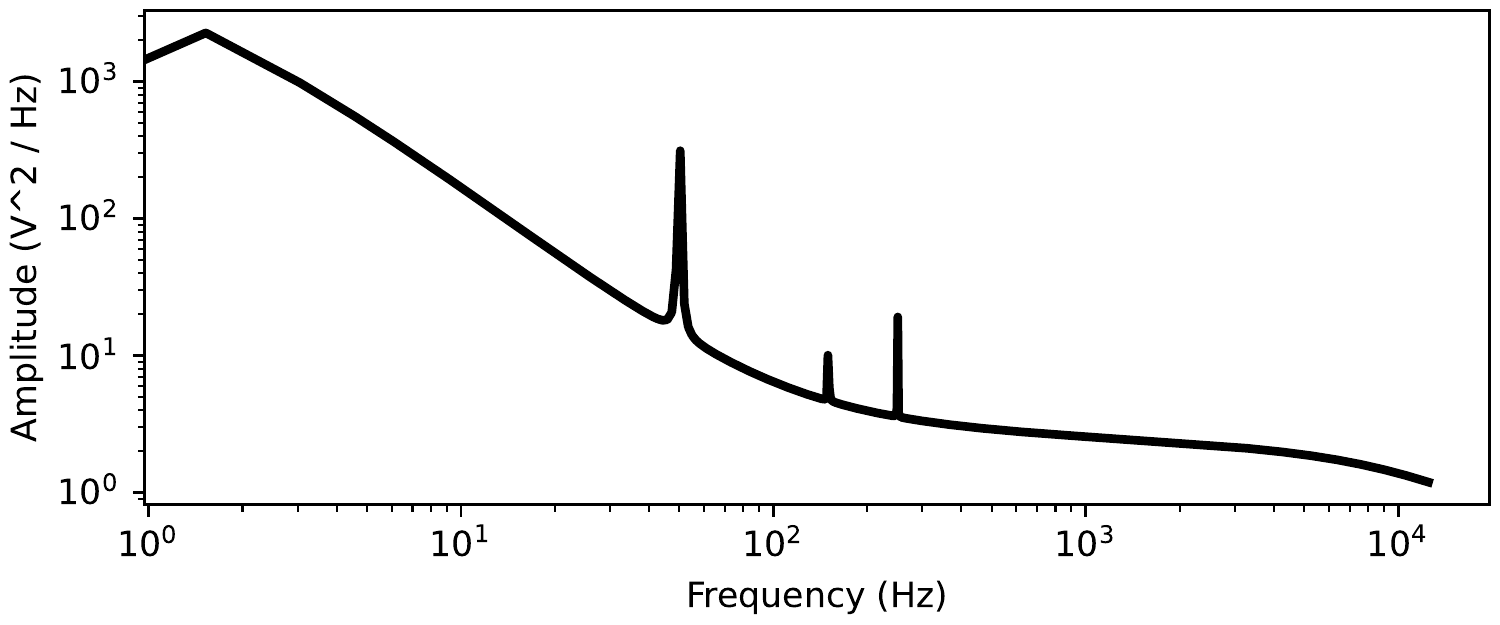}
\caption{Example of a simulated NPS.}
\label{nps}
\end{figure}

\begin{figure}[t]
\centering
\includegraphics[width=\linewidth]{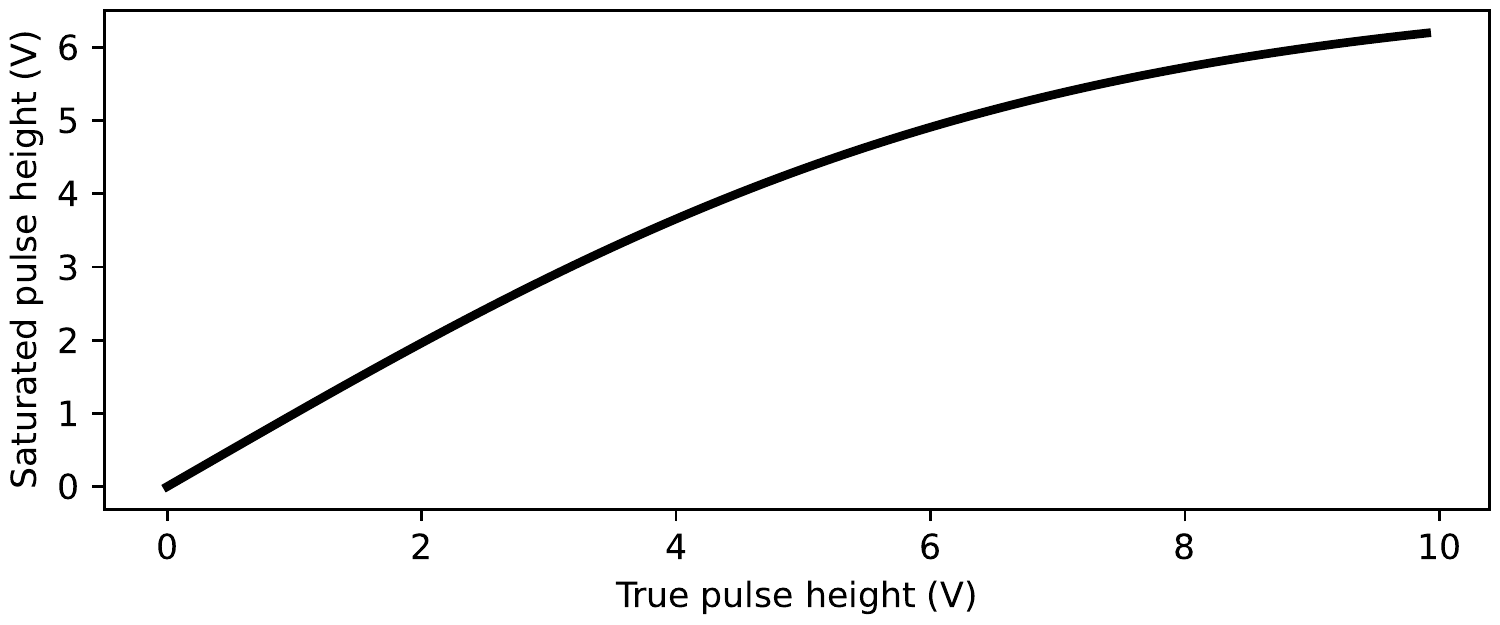}
\caption{Example of a simulated saturation curve.}
\label{sat}
\end{figure}

A target recoil and a set of typical artifacts were already shown in Fig.~\ref{sim_events}. A list of event classes which we simulate is shown and described in Box \ref{evtypes}. The time series of the event classes are simulated with parametric templates, mimicking the shape of realistic events. The templates are superposed with simulated noise traces, the simulation follows the method described in Ref. \cite{carrettoni_generation_2010}. In measurements, the noise of a detector is characteristic and well described by the NPS. An example of a simulated NPS is shown in Fig.~\ref{nps}. The shape of higher event pulses can be distorted by saturation effects of the TES. We approximate these saturation effects with a generalized logistics function, an example is shown in Fig.~\ref{sat}. 

For our experiments in the following section, we create a realistic data set from a virtual detector, with a total of 32500 events. The shares of the event types on the total size of the data set are listed in Tab.~\ref{tablesets} (test set). The noise baselines are simulated with a common NPS and a polynomial drift distribution, a saturation curve is applied to all events. Both target recoils and carrier recoils are simulated, i.e. two different types of pulse shapes appear in the data set. The record length is chosen as 16384 samples per time series, with a time base of 40 $\mu$s. 

Additionally, we create a second data set with more diversity in the event classes. This data set, consisting of 64000 events in total, we will use in the subsequent section to train a Convolutional Neural Network (CNN) for event classification \cite{726791}. 
The distribution of event types is again shown in Tab.~\ref{tablesets} (training set). There is one significant difference between the simulated training set and the test set: the pulse shapes, NPS and saturation curves in the training set are sampled on an event-by-event basis, i.e. each event features characteristics as if it originated from an individual measurement and detector. The distributions are set to cover the complete parameters space that can appear in realistic data.

To our knowledge, the simulation of multiclass data sets, including target recoils and artifacts, was not studied in the literature previously, nor implemented in software packages other than Cait. 

\begin{table}[t]
\centering
\begin{tabular}{lcc} 
\toprule
Event type & Test set & Training set \\ \midrule
Event pulse & 8000 & 20000 \\
Noise & 3000  & 10000  \\
Decaying baseline & 500  & 1500 \\
Temperature rise & 500  & 1500 \\
Spike & 500  & 1500 \\
SQUID jump & 500  & 1500 \\
Reset & 500  & 3000 \\
Tail event & 500  & 5000 \\
Decaying baseline & & \\
with event pulse & 500  & 5000 \\
Decaying baseline & & \\
with tail event & 500  & - \\
Pile up & 9000  & 10000 \\
Early or late trigger & 500  & 5000 \\
Carrier event & 8000  & - \\ \bottomrule
\end{tabular}
\caption{Event classes in the simulated data sets.}
\label{tablesets}
\end{table}

\subsection{Universal data cleaning}
\label{sec:universaleventselection}



A human expert can distinguish the time series of event types, regardless of the specific properties of the detector and without prior knowledge of the detector. It is therefore possible to generalize the unique shape of the event types across specific data sets. With the universal training set described in the previous section, we train a CNN for the binary classification task of discriminating target recoils from artifacts. 
    
The module architecture is intentionally kept simple and listed in Tab.~\ref{module}. The universal training set is split according to the ratio 7:2:1 into a training set, a validation set and a test set respectively. All time series are downsampled to a length of 512 in the preprocessing, and normalized such that the minimum and maximum of each individual time series are 0 and 1, respectively. The model is trained for 70 epochs, with an ADAM optimizer \cite{2015-kingma}, a learning rate of 0.001 and a batch size of 64. The negative log likelihood loss function is used as training objective. The training is done within the PyTorch Lightning framework. The full training takes 23 minutes on a 2 GHz Quad-Core 10th gen Intel Core i5. The final loss for the training set is 0.070, for the validation set 0.081, and for the test set 0.085. 

\begin{table}[t]
\centering
\begin{tabular}{p{2.5 cm} p{4 cm}} 
\toprule
Layer & Specifications \\ \midrule
1D convolutional layer & 1 input channel, 50 output channels, kernelsize 8, stride 4 \\
1D convolutional layer & 50 input channel, 10 output channels, kernelsize 8, stride 4 \\
Feed forward layer & 639 input nodes, 200 output nodes, ReLU activation function \\
Feed forward layer & 200 input nodes, 2 output nodes, ReLU activation function \\ \bottomrule
\end{tabular}
\caption{The CNN model architecture.}
\label{module}
\end{table}

We test the performance of our model on the realistic test set, which was described in the previous section. Overall we achieve an unweighted accuracy score of 0.941 across all event classes, an F score of 0.943, a precision of 0.902, and efficiency of 0.989. The survival rate of target recoils with low signal-to-noise ratio is visualized in Fig.~\ref{efficiency}. The good survival rate down to 5 times the noise resolution ($\sigma$), typical trigger threshold value, and the complete rejection of noise events above threshold makes the model suitable for low energy analysis. Above threshold, the survival rate of target events asymptotically approaches unity. The total survival rates of all event classes are shown in Tab.~\ref{surv_rates}. To interpret these scores, we have to consider several aspects:

\begin{figure}[t]
\centering
\includegraphics[width=\linewidth]{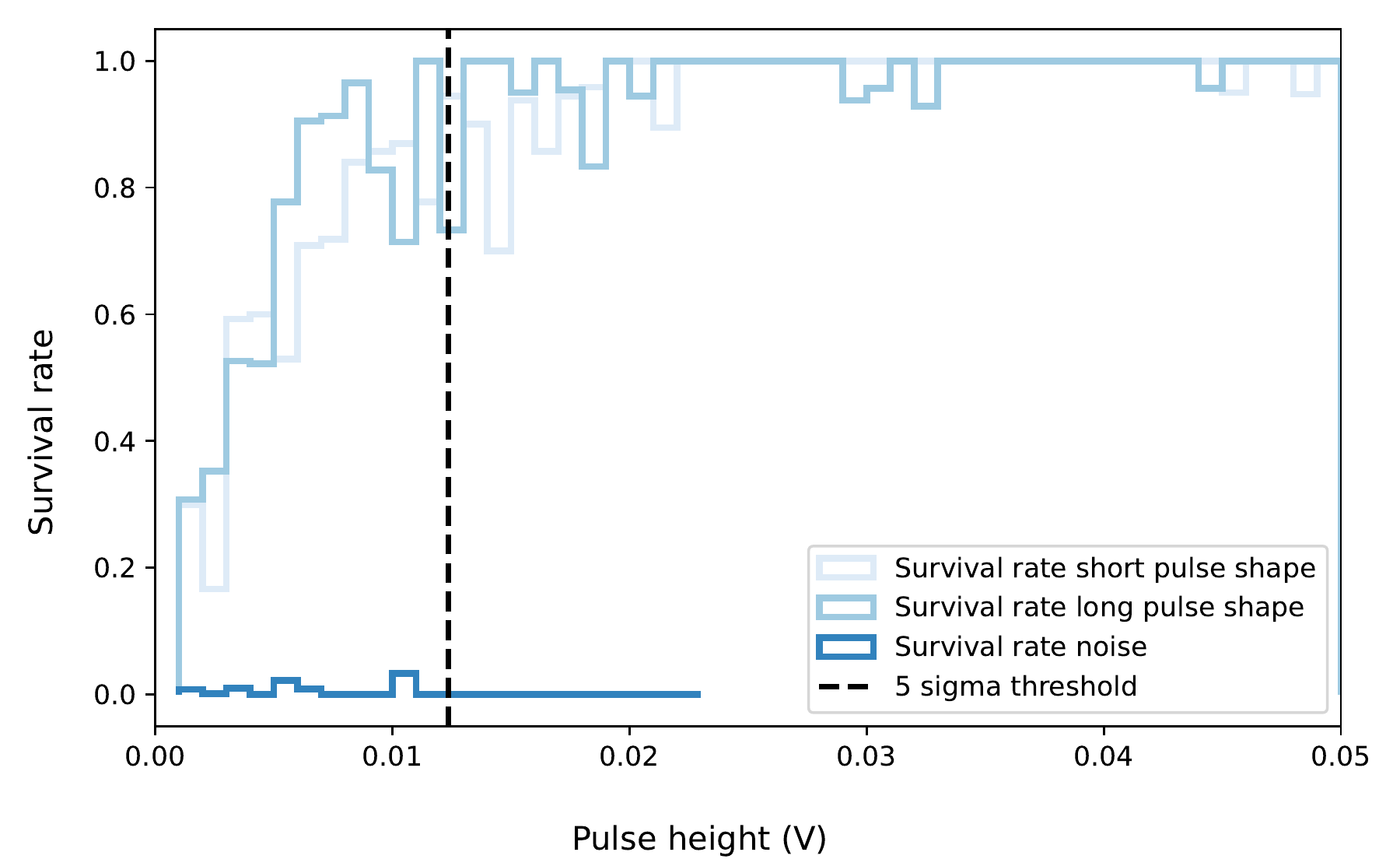}
\caption{Survival rate of different event classes after application of the CNN as a quality cut. As a reference, a typical threshold value is indicated by a black dashed line, at five times the value of the noise resolution $\sigma$. The long pulse shapes (medium blue) and the short pulse shapes (light blue) achieve both an efficiency around 0.9 down to the threshold. No noise events (dark blue) survive the CNN cut above threshold.}
\label{efficiency}
\end{figure}

\begin{figure*}[!h]
\centering
\includegraphics[width=\linewidth]{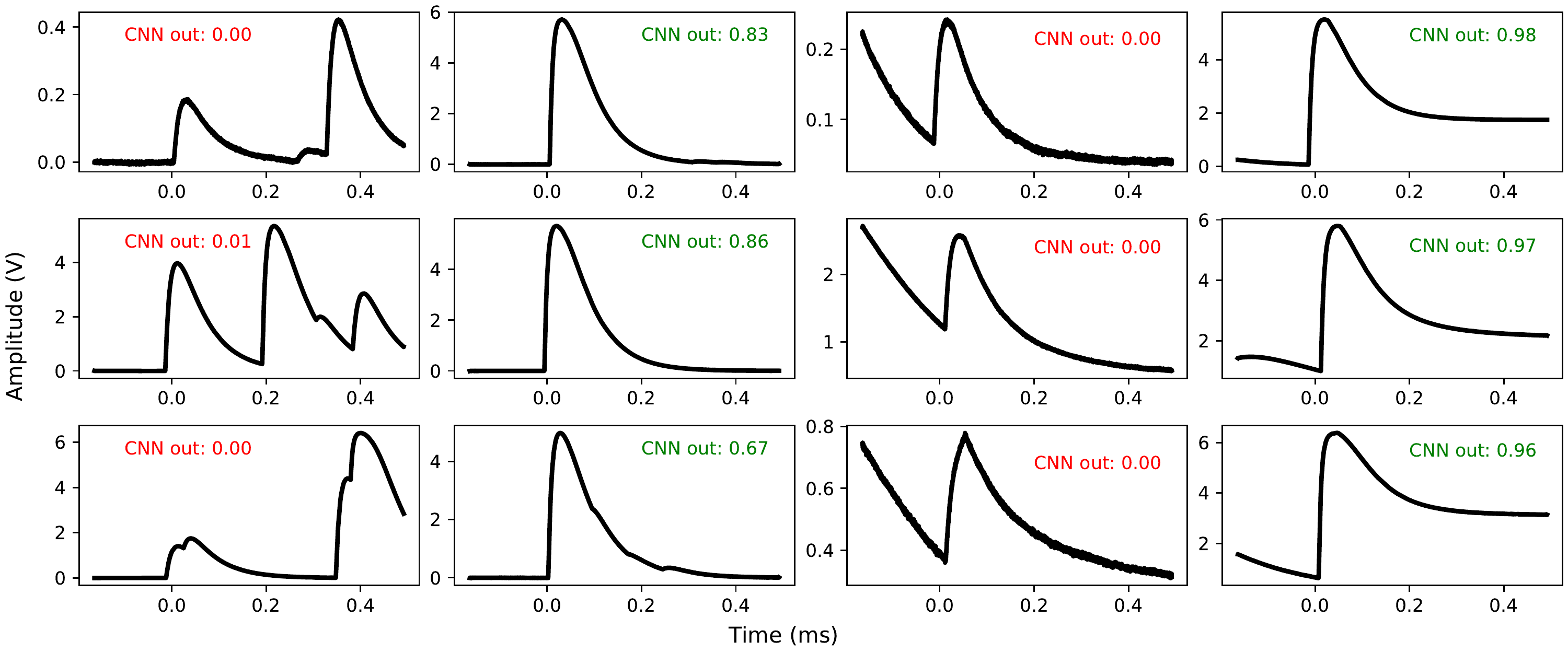}
\caption{Exemplary ``Pile Up'' (first and second columns) and ``Decaying baseline with tail event'' (third and fourth column) events from the test set. For the first event class, only pulse shapes with small piled up, secondary pulses survive the cut (CNN out $>$ 0.5), while severe pile ups are significantly rejected (CNN out $<<$ 0.5). The second event class was not included in the training set and serves therefore as a metric for the performance of the CNN on yet unseen event classes. Events with strong decaying baseline get rejected, while events with only weak decaying baseline survive the cut.}
\label{false_positive}
\end{figure*}

\begin{table}[b]
\centering
\begin{tabular}{p{3 cm} p{1.5 cm} p{1.5 cm}} 
\toprule
Event type & Number survived & Survival rate ($\%$) \\ \midrule
Event pulse & 7922/8000 & 99.0 \\
Noise & 17/3000 & 0.6 \\
Decaying baseline & 0/500 & 0 \\
Temperature rise & 0/500 & 0 \\
Spike & 7/500 & 1.4\\
SQUID jump & 3/500 & 0.6 \\
Reset & 5/500 & 1.0 \\
Tail event & 497/500 & 99.4 \\
Decaying baseline &  & \\
with event pulse & 0/500 & 0.0 \\
Decaying baseline &  & \\
with tail event & 107/500 & 21.4  \\
Pile up & 1088/8000 & 12.1 \\
Early or late trigger & 0/500 & 0 \\
Carrier event & 7896/8000 & 98.7 \\ \bottomrule
\end{tabular}
\caption{Survival rates of event classes in the test set.}
\label{surv_rates}
\end{table}

\begin{itemize}
    \item The event class ``Carrier Event'' is qualitatively similar to the class ``Event Pulse'', but with shorter pulse shape. As we sampled a wide range of pulse shapes for our training set, the scenario of carrier events is covered.
    \item The surviving ``Pile Up'' artifacts are mostly pulses with large height difference (see Fig.~\ref{false_positive}). More pile ups could be rejected with a lower acceptance threshold for the CNN output value (per default 0.5), or by additionally applying the peak detection used in Sec.~\ref{sec:detchar}. 
    \item The event class ``Decaying Baseline with Tail Event'' was not included in the training set, and therefore serves as a test of the generalization to previously unseen artifacts (see also Fig.~\ref{false_positive}).
\end{itemize}

Overall, our universally trained CNN model shows good performance as a multi-purpose tool for automated data cleaning on our test set. Our model has therefore a wide range of applications, e.g. for the large scale analysis of a multi detector setup, or the fully blind check of data quality. We stored the PyTorch Lightning checkpoint of the trained model in the resources of the Cait library. It can make predictions on a new data set without further training (code example below).

\vspace{0.5cm}
\begin{minted}[
baselinestretch=1,
% linenos,
fontsize=\footnotesize
]{python}
from cait.models import CNNModule
from cait.resources import get_resource_path

ckp = get_resource_path('cnn-clf-binary-v2.ckpt')
cnn = CNNModule.load_from_checkpoint(ckp)

ai.models.nn_predict(h5_path='./mock_001.h5',
           model=cnn,
           feature_channel=0)
\end{minted}
\vspace{0.5cm}

The prediction time of the CNN is of $\mathcal{O}$(10 k) events per second, significantly faster than the calculation of time series features or the application of fitting techniques on event-by-event basis. The method presented above is therefore a promising candidate for fast, universal and parameter-free data cleaning.

We want to emphasize that our method does not necessarily outperform a handcrafted, traditional cut analysis. We managed to create cuts for our test set that achieve an unweighted accuracy score of 0.959, better than the CNN. This result is expected: a cut analysis is designed on the data set that shall be cut (non-blind), or a subset dedicated for training (half-blind), while our method is applied fully blind, without manual human interventions for the individual detector.


\subsection{Interactive data visualization}
\label{sec:viztool}

\begin{figure*}[!t]
\centering
\subfloat{\includegraphics[width=0.9\linewidth]{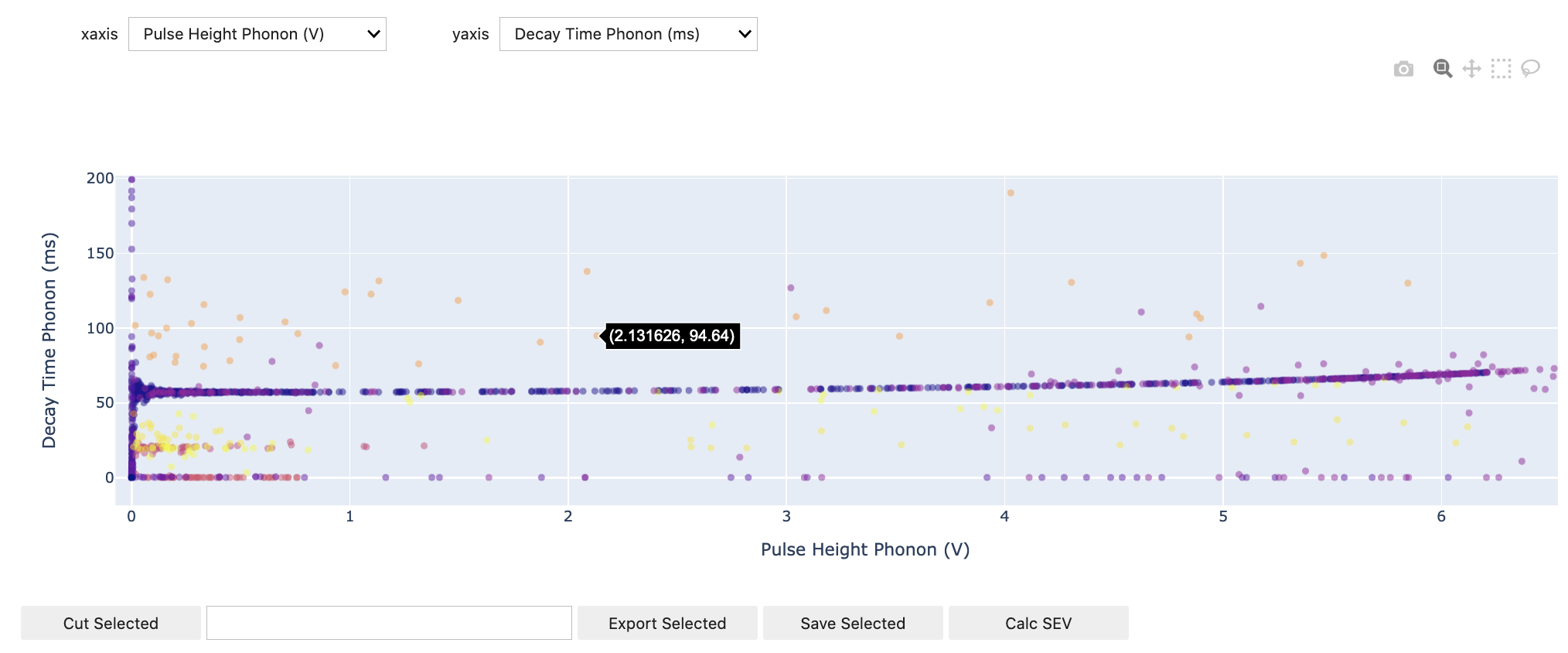}} 
\hfill
\subfloat{\includegraphics[width=0.9\linewidth]{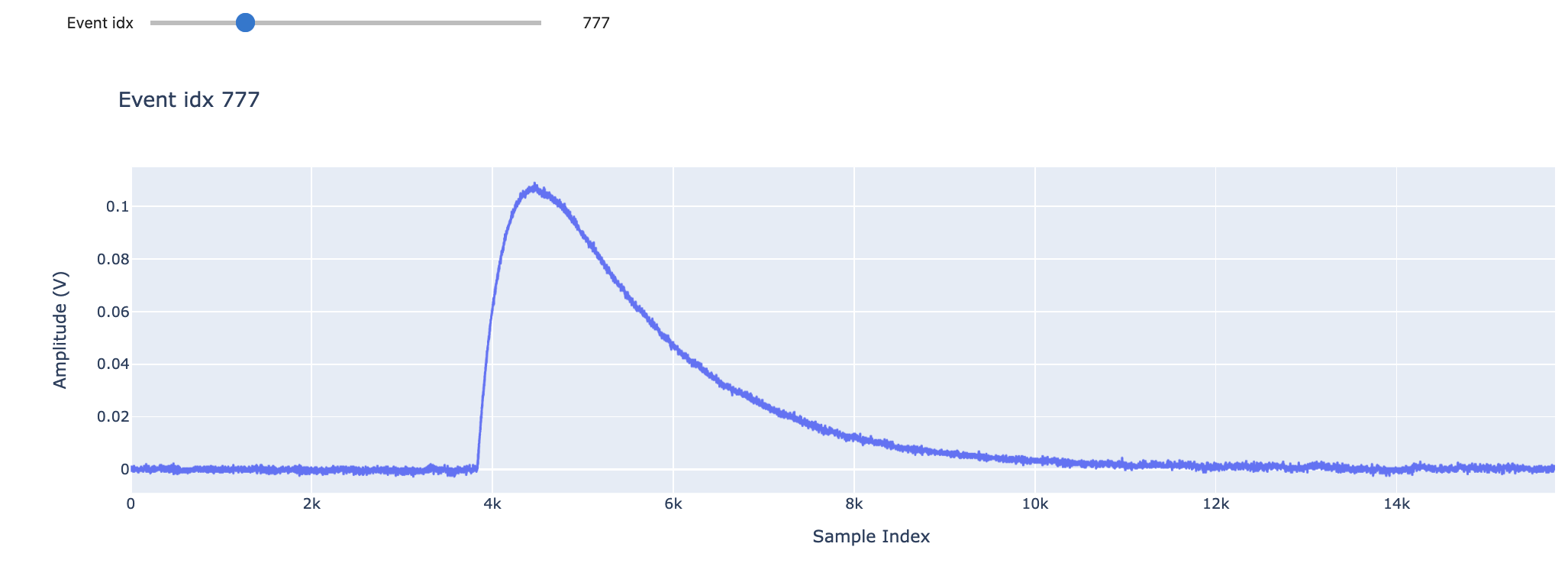}} 
\hfill
\subfloat{\includegraphics[width=\linewidth]{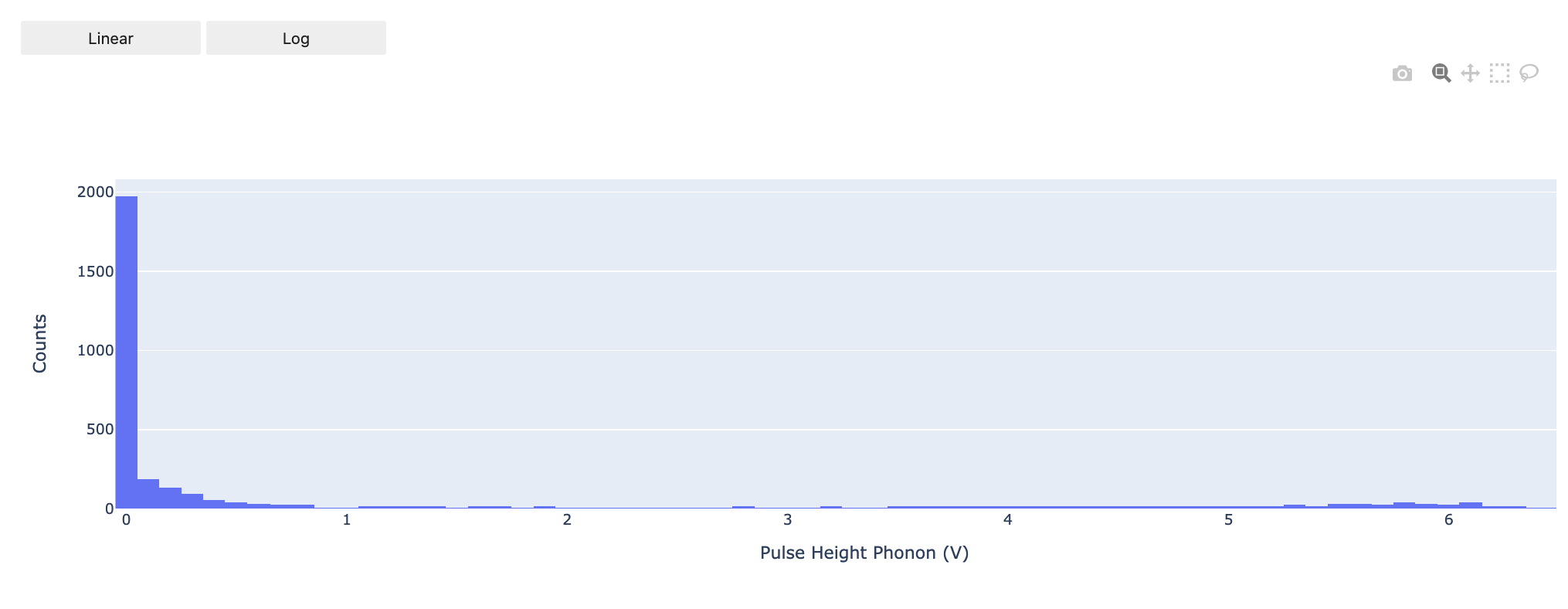}}
\caption{Screenshot of the visualization with a \textit{VizTool} instance. The code producing this output is contained in the documentation \cite{cait_rtd}. In the top row, two drop down menus let the user choose the data sets which are used as abscissa and ordinate for the scatter plot, displayed directly below. A click on the marker of an event in the scatter plot visualizes the event's time series in the center row. Selecting events on the scatter plot with the rectangular or lasso tool, produces an internally stored list of indices. With the \textit{Event idx} bar, we can scroll through these events. In the lowest row, a histogram of the selected events is displayed.}
\label{viztool}
\end{figure*}

The detector characterization and event selection steps of the analysis data workflow require an interactive setup: the analyst needs to decide on suitable cut values, based on one- or two-dimensional plots of parameter distributions. To gain insight into the nature of event clusters, exemplary time series of events are plotted individually. Cait comes with an interactive tool for this process (see Fig.~\ref{viztool}): in two drop down menus, a combination of features can be chosen to produce a scatter plot and individual histograms. Each data point on the scatter plot corresponds to an event, and its time series is accessible with a mouse click. A subset of events can be selected, to enable fast scrolling through the corresponding time series. Previews of standard events from selected events can be produced, and selections can be stored in the HDF5 file, and used as event cuts. A code example, using the test set from Sec.~\ref{sec:simulation} and producing the output shown in Fig.~\ref{viztool}, is shown below.

\vspace{0.5cm}
\begin{minted}[
baselinestretch=1,
% linenos,
fontsize=\footnotesize
]{python}
datasets = {
    'Pulse Height Phonon (V)': ['pulse_height', 0, None],
    'Rise Time Phonon (ms)': ['rise_time', 0, None],
    'Decay Time Phonon (ms)': ['decay_time', 0, None],
    'Onset Phonon (ms)': ['onset', 0, None],
    'Slope Phonon (V)': ['slope', 0, None],
    'Variance Phonon (V^2)': ['var', 0, None],
    'Mean Phonon (V)': ['mean', 0, None],
    'Skewness Phonon': ['skewness', 0, None],
}

viz = ai.VizTool(path_h5='./', 
              fname='mock_001',
              group='events', 
              datasets=datasets, 
              nmbr_channels=1, 
              sample_frequency=25000,
              record_length=16384)
              
viz.set_colors(color_flag=dh.get('events', 
'prediction')[0])

viz.show()
\end{minted}
\vspace{0.5cm}


The tool is designed for usage inside a Jupyter session \cite{Kluyver2016jupyter}. 

The typically instantaneous response time of the tool makes for an efficient analysis workflow, especially when compared to the subsequent execution of individual visualization commands. It enables fast identification of outliers and artifact clusters, and better choices of cuts. For new analysts, the interactive tool helps them gain a deeper understanding of the analysis process. While similar tools in stand-alone software exist, we are not aware of another one that can be fully integrated in the Python-based workflow.

\section{Conclusion}\label{sec:conclusion}

We presented Cait, a Python package for all tasks typically required for raw data analysis of cryogenic detectors. We review modern use cases covered by Cait, such as the triggering of continuously sampled data streams or the analysis of a multi-detector setup.
Triggered data is organized in HDF5 files, and is accessed and processed through the \verb|DataHandler| class. The workflow of a standard analysis of a detector has four steps: In the beginning, we characterize the detector with standard events, filters and values for the resolution and thresholds. In the second step, we trigger the full data stream with optimized thresholds and calculate features of individual events. In the event selection we establish decision rules, according to which events are either kept or discarded. In the last step, we validate our analysis with simulated events and estimate survival rates for target recoils. 

In the last chapter of this work, we highlighted three implemented methods. First, we simulate data sets that cover all typical classes of events, both target recoils and artifacts. Second, we train a CNN model to identify recoil events against artifacts and achieve promising accuracy and efficiency scores, readying our pre-trained model for fully automated data quality assessments. Third, we show an interactive Jupyter session for the event selection. 

Cait serves as an example of a modern analysis framework, based fully on Python. It includes implementations of all standard steps of the analysis, as well as experimental methods for a machine-learning based analysis. As an open source software package, it can be tailored to any project with individual extensions.

We expect that the performance of our CNN model presented in Sec.~\ref{sec:universaleventselection} can be improved by training and testing with larger data sets and ewith measured, not only simulated, data. At the time of writing this is the subject of ongoing work, in cooperation with the CRESST collaboration.


\backmatter

\bmhead{Acknowledgments}

We thank the CRESST and COSINUS collaborations for many discussions, especially Franz Pröbst, Martin Stahlberg, Nahuel Ferreiro Iachellini, Daniel Schmiedmayer and Christian Strandhagen. We are grateful for all early users, among them especially Rituparna Maji, who provided crucial feedback to the project. The computational results presented were obtained using the Vienna CLIP cluster. FW was supported by the Austrian Research Promotion Agency (FFG), project ML4CPD.  

\bmhead{Statements and Declarations}

On behalf of all authors, the corresponding author states that there is no conflict of interest. The datasets generated during and/or analysed during the current study are available from the corresponding author on reasonable request.










\begin{appendices}

\section{Code to simulate a mock data stream}
\label{app:mockstream}

\vspace{0.5cm}
\begin{minted}[
baselinestretch=1,
% linenos
]{python}
import cait as ai
import numpy as np

sample_frequency = 25000
duration = 2*3600
record_len = 16384
resolution = 0.01
wide = 0.01
nmbr_events = 200

stream = np.random.normal(loc=0, 
    scale=resolution, 
    size=duration*sample_frequency)
phs = np.random.normal(loc=1, 
    scale=wide, size=nmbr_events)
onsets = np.random.randint(
    low=record_len,
    high=stream.shape[0] - record_len, 
    size=nmbr_events)

t = np.arange(0, record_len, 1)
t /= sample_frequency
sev = ai.fit.pulse_template(t, 
    t0=record_len/sample_frequency/4, 
    An=1, At=0.5, 
    tau_n=0.01, tau_in=0.005, 
    tau_t=0.015)
sev /= np.max(sev)

for p,o in zip(phs, onsets):
    stream[o:o+record_len] += sev
    
stream = ai.data.convert_to_int(stream)
stream = stream.astype(np.int16)
stream.tofile('mock_001.csmpl')
\end{minted}
\vspace{0.5cm}




\end{appendices}


\bibliography{biblio}


\end{document}